# Multichannel Robot Speech Recognition Database: MChRSR

José Novoa, Juan Pablo Escudero, Josué Fredes, Jorge Wuth, Rodrigo Mahu
and Néstor Becerra Yoma

Speech Processing and Transmission Lab.
Universidad de Chile
Av. Tupper 2007, P.O.Box 412-3
Santiago-CHILE
E-mail: nbecerra@ing.uchile.cl     Tel: +56-2-29784205

## Abstract

In real human-robot interaction (HRI) scenarios, speech recognition represents a major challenge due to robot noise, background noise and time-varying acoustic channel. This document describes the procedure used to obtain the Multichannel Robot Speech Recognition Database (MChRSR). It is composed of 12 hours of multichannel evaluation data recorded in a real mobile HRI scenario. This database was recorded with a PR2 robot performing different translational and azimuthal movements. Accordingly, 16 evaluation sets were obtained re-recording the clean set of the Aurora-4 database in different movement conditions.

## 1. Database Recording

The experimental setup used in the database recording employs a PR2 robot which is a state-of-the-art mobile manipulation robot. It has a Microsoft Xbox 360 Kinect sensor mounted on the top. We re-record the clean test set from Aurora-4 database [1] in a meeting room considering different relative movements between the speech source and the robot. A TANNOY 501a loudspeaker was used as the audio source. The recording process was performed with the PR2's Microsoft Kinect sensor which has a four-microphone array. The re-recording was carried out while the robot was performing translational and head rotation movements simultaneously. Before the recording of each robot movement condition, the background noise was measured and the maximum of equivalent sound pressure level ($L_{eq}$) over ten minutes was 39dBA.

## 2. Robot Movements

During the recording of the different sets of the evaluation database, the robot made two types of movements that typically could be found in an HRI scenario: robot translation and robot's head azimuthal rotation.

### 2.1 Robot Translation

The first movement defined for the robot was the translational movement. Here, the robot moved towards and away from the loudspeaker between points P1 and P2 in Fig. 1. Three values for the robot displacement velocity were selected: 0.30 m/s, 0.45 m/s and 0.60 m/s. Those values were inspired by the discussions in [2], where a robot approached to a seated person at 0.25 m/s and 0.40 m/s. In those conditions, none of the human participants found these robot speeds too fast or disturbing. The selected velocities were multiplied by the speed factor function shown in Fig. 2. An additional test database was recorded with the robot at point P1. Thus, four robot displacement conditions were considered for the data recording: a static condition at position P1, and three translational movements between points P1 and P2.

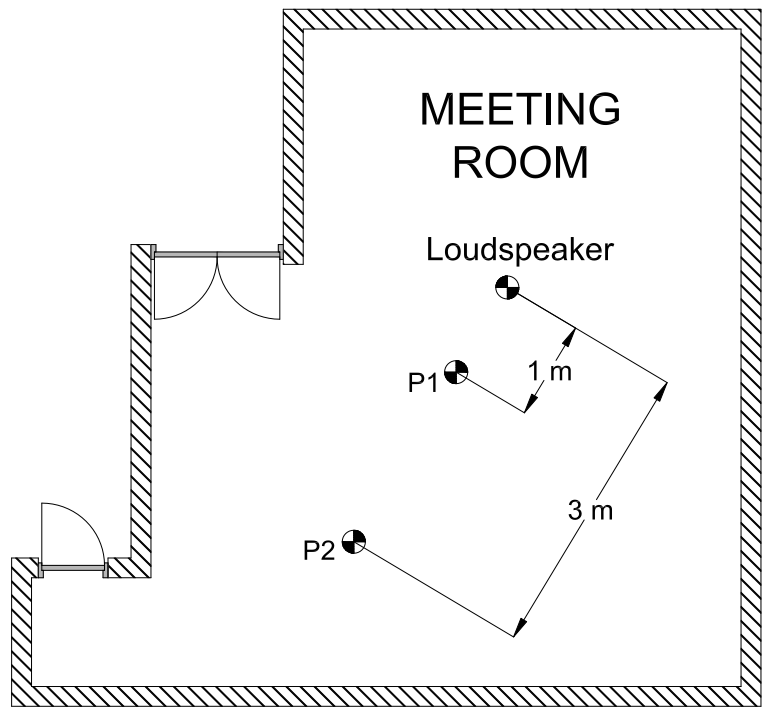

**Figure 1**: Recording diagram in the meeting room floor plan. In the static condition, i.e. without robot translation, the robot was positioned at P1. In the dynamic conditions the robot moved towards and away from the loudspeaker between points P1 and P2 which are located at 1 and 3 meters from the loudspeaker, respectively.

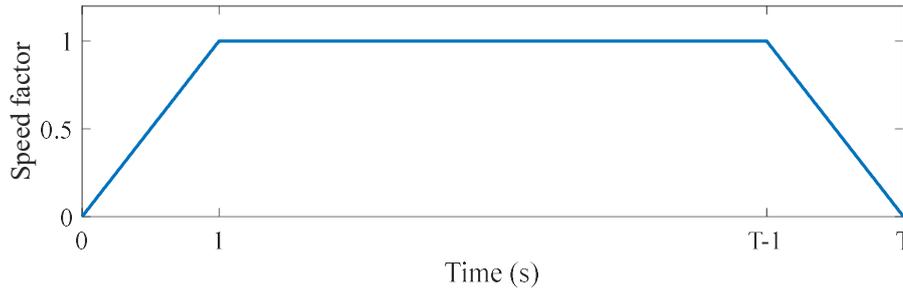

**Figure 2:** Speed factor function, T corresponds to the one-way travel time from point P1 to point P2 in the Fig. 1 or vice-verse.

## 2.2 Robot Head Azimuthal Rotation

The robot rotates its head making an azimuthal sweep, as can be seen in Fig. 3a. The robot's head moved periodically, sweeping between –150º and 150º, at three different angular velocities. During the movements, the front of the robot was towards to the loudspeaker which was situated at 0°. The three selected values for the angular velocity were: 0.28 rad/s, 0.42 rad/s and 0.56 rad/s. These values correspond to the angular velocity necessary to follow with the robot's head a virtual target. The virtual target was considered placed two meters from the robot and moving with a tangential velocity equal to 2 km/h, 3 km/h and 4 km/h, respectively, as shown in Fig. 3b. Additionally, a fourth angular robot´s head motion condition was generated by positioning the head oriented towards the loud speaker and making the head angular velocity equal to zero. Figure 4 shows the clean and the resulting spectrograms for a given utterance recorded with the Kinect microphones at the most severe motion condition, i.e. with angular and displacement velocities equal to 0.56 rad/s and 0.60 m/s, respectively.

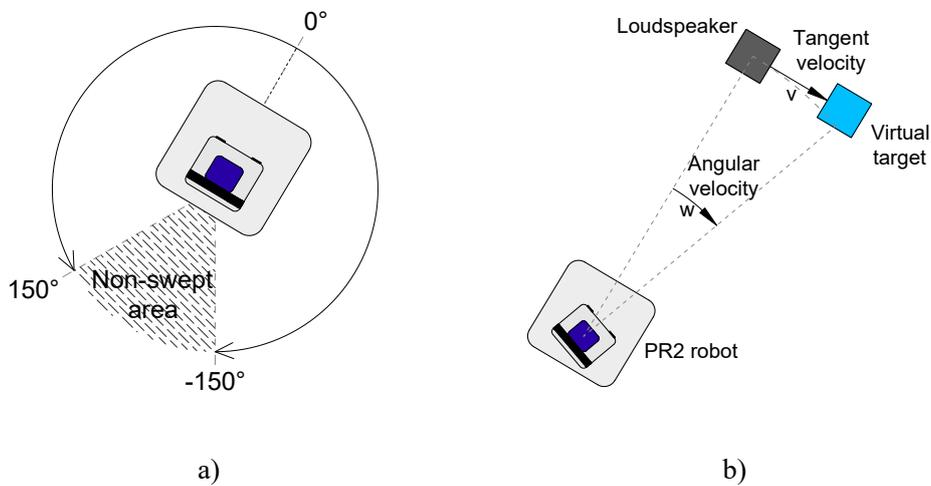

a)   b)

**Figure 3**: a) Angle swept by the azimuthal movement performed by the robot's head during the utterances recording. The loudspeaker was located at 0º. The robot's head moves periodically from -150º to 150º at different angular velocities. Recordings with static head were performed at 0º (i.e., oriented towards the source). b) The selected angular velocities of the robot's head movement correspond to the velocity necessary to follow with the head a virtual target located two meters away and moving with tangential velocities equal to 2 km/h, 3 km/h and 4 km/h.

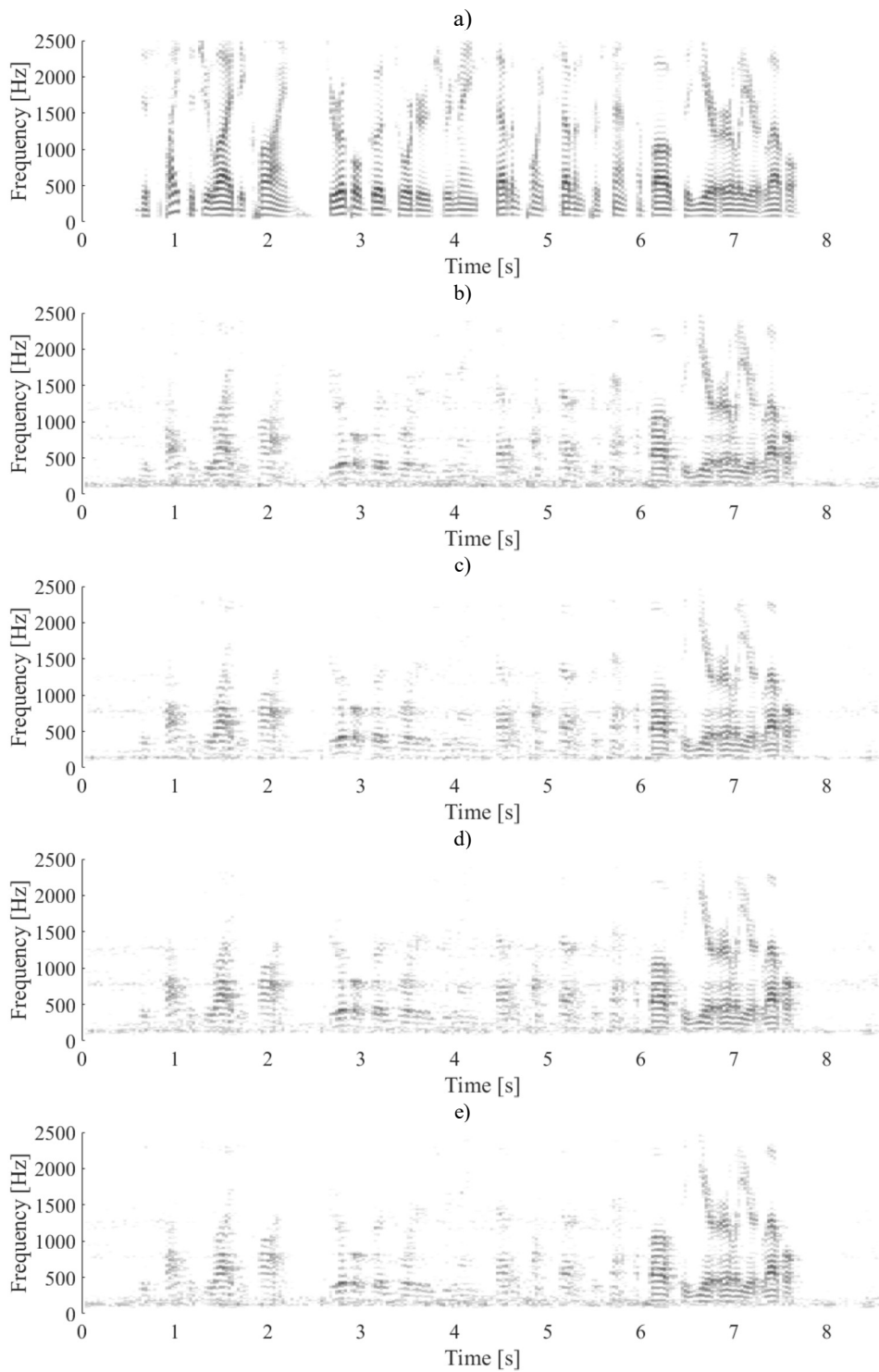

**Figure 4**: Original and resulting spectrograms for a given utterance recorded with the Kinect microphones: a) the original clean utterance; and, b), c), d) and e) correspond to the four Kinect microphones when the angular and displacement velocities were made equal to 0.56 rad/s and 0.60 m/s, respectively.

Summarizing, 16 evaluation sets were generated by re-recording the clean test sets of the Aurora-4 database in different translational and azimuthal motion conditions which correspond to 12 hours of multichannel evaluation data. The MChRSR data is available at http://www.lptv.cl/en/hri-asr/. More inquiries can be made directly to the last author of this paper.

## Acknowledgements

The research reported here was funded by Grants Conicyt-Fondecyt 1151306 and ONRG N°62909-17-1-2002. José Novoa was supported by Grant CONICYT-PCHA/Doctorado Nacional/2014-21140711.